\documentclass[12pt,preprint]{aastex}

\shorttitle{Three Dimensional ISM}
\shortauthors{Yeunjin Kim et al.}

\begin{document}

\title{Structure of the Interstellar Medium Around Cas A}

\author{Yeunjin Kim \altaffilmark{1,2},
G. H. Rieke \altaffilmark{1},
O. Krause \altaffilmark{3},
K. Misselt \altaffilmark{1},
R. Indebetouw \altaffilmark{4},
K. E. Johnson \altaffilmark{4}}

\altaffiltext{1}{Steward Observatory, University of Arizona, 933 North Cherry Avenue, 
Tucson, Arizona 85721}
\altaffiltext{2}{currently at the Department of Astronomy and Astrophysics, 
The University of Chicago, 5640 S. Ellis Ave., Chicago, Illinois 60637}
\altaffiltext{3}{Max-Planck-Institut f\"ur Astronomie, K\"onigstuhl 17, 69117 Heidelberg}
\altaffiltext{4}{Astronomy Department, University of Virginia, P. O. Box 3818, Charlottesville,
VA 22903-0818}

\begin{abstract}

We present a three-year series of observations at 24$\mu$m with the 
{\it Spitzer} Space Telescope of the 
interstellar material in a 200 x 200 arcmin square area centered on Cassiopeia A. 
Interstellar dust heated by the outward light pulse 
from the supernova explosion emits in the form of compact, moving features. Their 
sequential outward movements allow us to study the 
complicated {\it three-dimensional} structure of the interstellar medium (ISM) behind 
and near Cassiopeia A. The ISM consists of sheets and filaments, with 
many structures on a scale of a parsec or less. The spatial power spectrum 
of the ISM appears to be similar to 
that of fractals with a spectral index of 3.5. The filling factor for the 
small structures above the spatial wavenumber \textit{k} $\sim $ 0.5 cycles 
pc$^{-1}$ is only $\sim $ 0.4\%.

\end{abstract}


\section{Introduction}

The composition and structure of the interstellar medium (ISM) are critical 
to our understanding of the evolution of stars and galaxies. The gas is shaped by 
turbulence and magnetic fields and is highly structured (Verschuur 
1995). Instabilities cause the gas to collect into compact filaments and 
sheets that collapse into stars (e.g., V\'azquez-Semadeni et al. 2007 and 
references therein; Boss 2005). The interstellar grains account for 
only 1\% of the total mass of the ISM, but their interaction with stellar
photons is the dominant source of opacity and hence strongly influences
star formation and evolution. There have been several 
observations that give us some clues to the nature of the dust in the
ISM: for example, the behavior of Polycyclic Aromatic Hydrocarbons (PAHs) in 
nearby galaxies (Smith et al. 2007), and the mass or size distribution of dust 
particles in the ISM (Kim \& Martin 1996; Mathis et al. 1977), to just name 
a few (see review by Draine 2003). However, as yet, no fully developed model 
has been developed for the detailed morphology of the ISM. 

An understanding of the ISM structure is necessary to understand
the forces and mechanisms that shape it and set the preconditions for
star formation and other widely studied processes (e.g., Carlqvist
\& Gahm 1992; Verschuur 1995; Fiege \& Pudritz 2000). The ISM structure is also 
a key input to studies of radiative transfer, and therefore is critical
to understanding the properties of dust-embedded sources (e.g., Bruzual 
et al. 1988; Witt \& Gordon 1996; Wolf et al. 1999; Hegmann \& Kegel 2003). 
Specific examples include young stellar object clumpy 
circumstellar media (e.g., Indebetouw et al. 2006), HII regions (e.g., Wood et al. 2005),
and galaxy disks (e.g., Inoue 2005) and references in these works. It has
been shown that the assumed structure can affect the wavelength dependence
of the attenuation (e.g., Fischera \& Dopita 2005) and
can significantly modify the observed spectral energy distribution of a source
(Indebetouw et al. 2006). Clumping can also allow self-shielding of the ISM
gas and therefore can alter observed line strengths (e.g. Burton et al. 1990).

The structure of the ISM is conventionally measured using 
HI observations. Verschuur (1995) reviews these results, emphasizing the 
prevalence of filamentary structures and the possible causes for them. HI 
data can identify such features through coherent velocity structures, but 
it is difficult to relate them accurately to one another along the line of 
sight. Braun \& Kanekar (2005) have recently used HI measurements to 
discover tiny interstellar clouds in the local ISM, with dimensions of 0.1 
pc or even less. There have been several efforts to study the fractal 
structures of the ISM (Elmegreen \& Falgarone 1996) 
on subparsec scales (Ingalls et al. 2004; Falgarone et 
al. 2004, and references therein). Small scale structures have also
been probed by observing the scintillation of signals from compact
radio sources, indicating that the density fluctuation spectrum is
consistent with Kolmorgov turbulence (e.g., Stinebring et al. 2000). 
More closely related to our study are 
three-dimensional maps of the dust structures in the foreground of supernova 
(SN) 1987A (Xu et al. 1995) and of type Ia SN in general (Patat 2005). 
These studies are based on the highly complex physical characteristics of 
interstellar dust particles, such as three-dimensional multiple scattering 
or polarization, with dependencies on both the dust parameters and the 
geometry. Thus, they must introduce additional parameters that are difficult 
to constrain and can increase the uncertainties in the results. 

Here we present a simple and straightforward method to study the structure 
of the ISM surrounding the supernova remnant (SNR) Cassiopeia A 
(Cas A) at a resolution
of $\sim$ 0.1pc. Cas A exploded 
approximately 10,000 years ago, and the light coming directly from the explosion 
reached Earth about 325 years ago (Thorstensen et al. 
2001). In 
previous work (Krause et al. 2005), we discovered infrared (IR) features at 
24$\mu$m up to about 25 arcmin from the center of the SNR. These features were 
apparently moving at superluminal velocities, much faster than the velocity 
of fast-moving knots around the SNR. Although initially some features appeared to
be associated with a more recent event in the core of the SNR, further tracking
of their motions now makes this explanation appear unlikely. We believe that all the observed IR 
features are the echoed reradiation of thermal energy 
as the light from the Cas A supernova 
passes through its surrounding ISM and heats the dust. Because of the indirect
path to reach us and the resulting increased propagation time, we see these
features now. These IR-radiating features, 
called IR echoes, are a novel approach to explore the distribution and the 
morphological structure of the ISM in three dimensions, 
and thus they solve the limitations of the previous integrated measures of the 
ISM structure in two dimensional, projected images. In this paper, we describe
the geometry of the echoes in \S 2, present observations of them over nearly
a three year interval in \S 3, develop and apply the tools to analyze the
data and construct images of the ISM in \S 4, discuss the results in \S5, and
summarize our conclusions in \S 6.

\section{Background}

IR echoes arise when the spherical pulse of electromagnetic radiation from 
the SN explosion encounters and heats interstellar dust. The resulting IR 
radiation traces the ISM along the radiation pulse. The IR echoes we observe 
at a given time travel over the same total distance from Cas A to dust 
grains in the ISM and then to the Earth, which requires that the echoes lie 
on an ellipsoid whose foci are Cas A and the Earth. Thus in a sequence of 
observations, the IR echoes must move successively outward to more distant
ellipsoidal regions of the ISM. We then view continuously changing circular 
cross-sections of the ellipsoids as the light passes through
interstellar space. 

By assuming that the distance to the SNR from the Earth is much greater than 
the distance between the observed echo and the source, the ellipsoidal 
structure can be simplified to a paraboloid (see Fig. 1.). The paraboloid
that forms the locus for the echoes at any time is defined by the set of
points such that the light travel distance from the supernove to the vertex
and then back to the observer, D + ct, equals the light travel distance to
any point on the paraboloid (light travel distance r), and then to the observer
(light travel distance D-z).

\begin{equation}
\label{eq1}
{D + ct = \left( {D - z} \right) + r},
\end{equation}

\begin{equation}
\label{eq2}
{h^{2} + z^{2} = r^{2}},
\end{equation}

\begin{equation}
\label{eq3}
{z = \frac{{1}}{{2}}\left( {\frac{{h^{2}}}{{ct}} - ct} \right)},
\end{equation}

\noindent
where \textit{D} is the distance between the SNR and the Earth, \textit{z} 
is the distance of the observed IR echo from the SNR with the positive 
direction toward the observer, \textit{r} is the distance from the 
SNR to the heated ISM dust (Fig. 1 demonstrates that the echoes seen
at any one time have a range of distances from the SNR), 
\textit{h} is the perpendicular distance of the echo to the 
line of sight between Cas A and the Earth, and \textit{t} represents the 
time difference between the observed outburst of Cas A and the echo observation date.

\section{Observations}

Between 2003 November 30 and 2006 October 2, we used the Multiband Imaging 
Photometer for {\it Spitzer} (MIPS; Rieke et al. 2004) to obtain 11 images at 24$\mu$m of Cas A in six epochs
with an interval of about 6 months between epochs. Figure 2 shows one such image. 
The first 5 images (IOC Campaign W,
PID 231, PID 233, PID 718, PID 20381) were centered on Cas A and observed in scanning 
mode, while the other 6, all at a single epoch, were taken in photometry 
mode (PID 30571), concentrating on a number of selected regions. The first two images 
(epochs 1 and 2, 2003 November 30 and 2004 December 2) were taken within a 
region about 52 arc min long by 12 arc 
min wide, the third image (epoch 3, 2005 February 3) about 120 arc min long by 120 arc min 
wide, the next two images (epochs 4 and 5, 2005 September 1 and 2006 February 14) about 200 
arc min long by 200 arc 
min wide; the last six images (epoch 6, 2006 October 2) are each about 20 arc min long and 20 arc min wide. 

\section{Data Processing and Analysis}

Initial reductions of these data used standard procedures 
centered around the MIPS Instrument
team Data Analysis Tool (DAT) (Gordon et al. 2005). 
Fine-scale moving features in the reduced images 
behave as predicted for IR light echoes in \S 2. For example, 
the most distant such features are at a projected distance of 
about 100 pc from the site of the SN, and many of them have arc-like 
morphologies pointing back toward the SNR. 

The reconstruction of the three-dimensional structure of the ISM from the IR echoes 
requires a number of additional data analysis steps to isolate 
the IR echoes from other features. They were performed in the 
following order: fast Fourier transform (FFT) filtering (\S 4.1), star cleaning (\S 4.2), 
subtraction of images (\S 4.3), and finally determination of the 
three-dimensional structure of the ISM (\S 4.4). 

\subsection{Fast Fourier Transform (FFT) Filtering}

In addition to the fine filamentary structures, the IR images include strong
extended emission. The extended 
structures resemble the diffuse galactic cirrus emission and appear to be 
more than 40 arcmin in scale. In comparison, the fine structures due to infrared echoes stand
out distinctly against the diffuse sky background but are fainter than many 
stars and nebular components. Due to the characteristic 
geometry (see Fig. 1), the motions of the IR echoes projected close to the
position of Cas A are fore-shortened and hence those echoes are generally 
observed to move more slowly than the echoes far out near the edges of our images. The typical estimated 
apparent motion near Cas A is about 2 arc sec within a year compared with 
about 18 arc sec at a position perpendicular to the center of Cas A relative to 
the line of sight. The bottom four panels in Fig. 2 show the changes in 
morphology of some of the IR echoes for about one and a half years of 
observations from epoch 3 to epoch 6. 

The echoes can be isolated by taking differences of images at different epochs. 
Because of the limited time interval 
covered by our observations, such images can reveal stucture down to spatial 
frequencies of about 1 cycle pc$^{-1}$ (since the motion of the supernova light pulse
is approximately 0.3 pc yr$^{-1}$). 
However, simply taking difference images between two consecutive epochs is 
not the optimum way to eliminate the sky background underneath or in front of the 
IR echoes. As the {\it Spitzer} telescope moves in space, it looks through various 
zodiacal cloud columns depending on the observing date, changing the level 
of zodiacal foreground emission. In addition, the 
echo features do not necessarily move far enough over our measurement period 
for us to measure the structure of the fixed background emission behind 
them. Therefore, low frequency background structure can distort our 
measurements of the echoes. 

To reduce these two effects, we use high pass 
filtering to suppress the spatial frequencies where the information on the 
ISM is compromised by our limited time range of observation anyway. To set 
the filter parameters, we took a difference image of subregions of epoch 3 and epoch 5. We 
selected these epochs to have large areal coverage (20 arcmin square) over
a reasonably long time baseline (1 year). We 
computed the radial power spectrum for the difference image by azimuthally 
averaging in equally spaced wavenumber bins. Fig. 3 shows a shaded surface 
representation of the logarithm of the power spectrum of the difference 
image on the left and the averaged power spectrum on the right. As expected, 
the power spectrum rolls over below 1.5 cycles pc$^{-1}$, which corresponds 
to the lowest frequency at which echoes observed over a 1 year period  
should produce signals in the difference image. Since our total observation
set covers three years, we set the filtering to $\sim $ 0.5 cycles pc$^{-1}$, 
so we can insure almost 
100\% transmission of the echo information (See Fig. 3). 

For computational convenience, each epoch image was disaggregated to form 
small square subregions of 1001 x 1001 pixels, avoiding the bright emission
within about 8 arcmin of Cas A. 
To eliminate the diffuse sky background emission, the high-pass 
filter was then applied to each small region with a cutoff frequency of 0.01 cycles per pixel, 
or 10 cycles per image width, corresponding to 0.5 cycles pc$^{-1}$. The cutoff 
frequency is where the power transmitted by the high pass 
filter reaches its half maximum (see Fig. 3). The filter was implemented 
using the FFT routine in Interactive Data Language (IDL) and applying a 
Butterworth high pass filter. The equation for this filter is

\begin{equation}
\label{eq4}
{high\_pass = 1.0 / (1.0 + ((cutoff / 
\textit{distfun})^{(2*order)})},
\end{equation}

\noindent
where \textit{cutoff} is the predetermined value meant to represent the 
threshold frequency, 0.01 pixel$^{-1}$, above which the filter passes information, the 
\textit{order} parameter determines the steepness of the filter; the higher 
the order, the steeper the band, and \textit{distfun} is an array of the 
high pass filter image before any adjustment is made. The value of 
\textit{order} was set to 1 to avoid possible artifacts due to the effect of 
the high filter \textit{order} on the frequency response. The 
\textit{cutoff} frequency was kept consistent in all images to retain 
properly scaled images for subtraction later (\S 4.3).

With the application of this filter, almost all of the diffuse background 
emission was satisfactorily removed, only leaving spatially varying sources 
such as IR echoes, plus stars and bright nebulosity.

\subsection{Star Cleaning}

Depending on when the observations were scheduled, the position angle (PA) of 
the {\it Spitzer} telescope may vary. At the Ecliptic latitude of $\sim $54$^{\rm o}$ for 
Cas A, there is a change in PA by between 0.5$^{\rm o}$ and 1$^{\rm o}$ per day. 
As a result, the relation of a star to its latent image and other artifacts changes. This 
effect requires that we remove point sources before subtraction of the 
images, thus minimizing any possible star residuals. 

Stars were located by a set of IDL DAOPHOT procedures. The 
first procedure of the set, \textit{find,} detected stars that passed the 
user-given criteria of FWHM= 4.0 and threshold intensity =1.0 in pixel 
values. We ran the \textit{find} routine with different trials of FWHM 
ranging from 3.0 to 6.0 until we settled on 4.0 as the typical star FWHM 
that detected most of the stars in our images. The threshold intensity was 
set to 1.0; the stars whose central intensity was less than 1.0 were 
suppressed close to zero in difference images (\S 4.3). 
After the stars 
had been located, aperture photometry was used to measure the magnitude of 
each star and the background. Several reference stars in each 
image were manually selected to define a point-spread function (PSF) and 
then the best estimated PSF was subtracted from every detected stellar image. 

Although most of the stars were eliminated by this method, some star 
residuals were still present, especially for the extremely bright stars 
whose visible images extend to more than 10 pixels. These bright stars were interactively 
suppressed to a low pixel value = 0.1 by subtracting the PSF, and later they were removed almost 
completely by image subtraction (\S 4.3). The scanning process with MIPS 
produced a series of quasi-star small dots from latent images of bright 
stars along the scanning direction. In each observing period, scanning was 
performed in a different direction, so using the same star position for all 
the different epochs did not remove the latent images. Thus, we created 
separate star position lists and applied them for those images that had the 
problem. Another artifact due to the bright point sources was vertical 
pinstripings, also known as jailbars, as a result of pixel saturation. Faint 
jailbars were cleared up by subtraction (\S 4.3), but interactive cleaning was 
performed for strong jailbars, based on local averages of the repetitive pattern. 

\subsection{Subtraction}


Once a light pulse is absorbed by dust grains, the IR light echoes are 
re-emitted quickly into successive paraboloids and can be identified via a 
change in either the apparent motion or pixel values. Especially for those 
whose changes in apparent position are too small to be seen by eye - for 
example the echoes coming from the same radial direction at different 
times - the pixel intensity change can be the better way to probe them. 

To implement this approach, we performed the following steps. First, the 
below-zero pixels in star-cleaned images were scaled to zero; we are only 
interested in the positive signals, and removing any negative pixel points 
will prevent any misinterpretation of the results after subtracting images. 
Then, two images that covered the same region but were taken on different 
dates were subtracted from each other. Echoes in the first image will
appear as positive pixel values, while those in the subtracted image will yield negative pixel 
values (see Fig. 4). However, there may also be residuals due to the 
fluctuation of intensity in the background level itself due to the 
varying position of {\it Spitzer} in its orbit. 
To deal with this possibility, we took a difference image of a 
region far away from any apparent echo structures in the two epochs. 
Several regions were tested, and the maximum 
difference-pixel residuals for each trial region were measured and found on average to 
be approximately within $\pm$ 0.7 in units of pixel values. Therefore, any points 
whose pixel values in the difference image are between $\pm$ 0.7 were 
considered as possibly coming from the diffuse background and rejected. 
Most of the low-pixel-valued echoes were then recovered; 
this approach enabled us to also trace echoes lying above/below the 
nebulosity. 

\subsection{Three-dimensional Structure of the ISM}

To produce three-dimensional maps of the ISM, we first set the center of Cas 
A to the reference point (0,0,0) in a three dimensional space with positive 
\textit{x} toward the west of Cas A, positive \textit{y} toward the north, 
and positive \textit{z} toward the observer. We computed the two 
dimensional projected coordinates, \textit{x} and \textit{y}, of the echoes 
that pass the threshold, $\pm$ 0.7 (\S 4.3). With the resulting \textit{x 
}and \textit{y} values, the projected distance from Cas A to the echoes in 
the plane of the sky,\textit{ h} in Eq. (\ref{eq2}), was calculated using the Pythagorean theorem for every point. 

We adopted the explosion date of Cas A to be A.D. 1671.3 $\pm $ 0.9 
(Thorstensen et al. 2001). Solving for the spatial distance, 
\textit{r} in Eq. (\ref{eq1}) and substituting it 
into Eq. (\ref{eq2}) gives the line-of-sight depth, \textit{z} in Eq. (\ref{eq3}).
We recorded the echoes using the 3-D plot procedure, XPLOT3D, 
in IDL. This procedure takes three vectors in \textit{x}, \textit{y,} and 
\textit{z} coordinates and plots them as scattered data points in three 
dimensions. Fig. 5 illustrates this process for images from the first
two epochs of our observational series. Fig. 6 shows the final three-dimensional image of echoes from 
all six epochs. The images are constructed in pointillist style, consisting 
of dots representing each position where
an echo was detected. No attempt is made to encode the intensity of
the echoes. To distinguish data points taken on different observing 
dates, different symbol colors are used.

\section{Discussion}
  
In this section, we first show that the energy pulse from the SN explosion 
has overtaken residuals from the progenitor massive star and its episodes 
of mass loss. As a result, the IR echoes reveal the structure of undisturbed 
ISM, which we then discuss.

\subsection{Three Possible Sources of IR-Emitting Dust}

We consider three possible origins for the dust probed by the IR light 
echoes: newly condensed SN ejecta, circumstellar material (CSM) produced by 
stellar winds as the progenitor star of Cas A evolved until its death by its 
supernova explosion, and the pre-existent ISM dust. 

High-velocity SN ejecta from Cas A have previously been identified with 
expansion speeds up to nearly 12,000 km~s$^{-1} $(Fesen 2001). Since the IR 
echoes lie well outside even the most rapidly moving ejecta, at positions 
consistent with their moving outward at the speed of light, we 
can exclude that they are associated with such ejecta.

We have also considered the possibility that the echoes 
could arise from the circumstellar dust released during stellar evolutionary phases. 
Garc\'ia-Segura et al. (1996) showed that the progenitor star of Cas A 
underwent the evolutionary sequence O-star, Red-Supergiant (RSG) star, and 
Wolf-Rayet (WR) star roughly 10$^{4}$ yr before its supernova explosion; they 
constrained its initial mass to be between 25\textit{M}$_{\odot}$ and 
35\textit{M}$_{\odot}$.

In a 35\textit{M}$_{\odot}$ star model, the wind velocity is estimated at a few 1000 
km~s$^{-1}$ both in the main sequence (MS) and WR stages, and falls to $<$ 100 
km~s$^{-1} $in the RSG stage (Garc\'ia-Segura et al. 1996; Dwarkadas 2004). About 
70\% of the total material loss in the wind from MS to WR stages is lost in 
the RSG stage. With a small wind velocity, a high wind density, and a low 
temperature, molecule and dust formation is likely in the RSG stage, but the 
RSG thin shell expands only up to $<$ 10 pc, 
due to its relatively slow wind velocity ($\sim $ 75 km~s$^{-1}$). 
Thus, the main contributor to pushing the high density RSG materials 
outwards from the center of the star is the strong WR wind with its wind 
velocity $\sim $ 1000 km~s$^{-1}$ lasting 10$^{4}$ yrs. The study of 
pulsations in RSG with high L/M ratio by Heger et al. (1997) also concluded 
pre-supernova stars go through a phase of violent pulsational instability 
several 10$^{4} $yr before the supernova explosion, which might add to the 
WR wind.

The upper limit for a mass loss shell boundary radius has been estimated at 
$\sim $ 38 pc for an ambient ISM density of 20 cm$^{-3}$ (Garc\'ia-Segura et 
al. 1996) or at $\sim $ 74 pc for an ISM density of 1 cm$^{-3}$ (Dwarkadas 
2004). To understand whether this boundary lies inside or outside of the IR 
echoes requires that we estimate the ISM density around Cas A. Krause et al. 
(2004) find a typical column density in gas of $\sim $ 40 mg cm$^{-2}$ in 
front of the SNR. Assuming this gas is distributed uniformly between us and 
the SNR would result in an estimate of $\sim $ 2 cm$^{-3}$ for the average 
ISM density around Cas A. However, the line of sight to the SNR intersects 
the Perseus Arm, the outermost main spiral arm of the Milky Way, and most of 
the material should also lie in this arm rather than in the space between it 
and the sun (Wilson et al. 1993). A typical width of a Milky Way arm is 1.32 
kpc (Russeil 2003). We can therefore assume that most of the interstellar 
material lies within 1 kpc of the SNR, compared with its distance of 3.4 
kpc. Thus, a plausible estimate for the ISM density in the space around Cas 
A is $\sim $ 6 cm$^{-3}$. Consequently, for average conditions, the boundary 
of the shell should lie at or within the 50 pc radius where the vertex of 
the paraboloid corresponding to epoch 1 is located and the IR echoes are 
observed at the shortest distance since the explosion of Cas A. In most 
locations, this argument indicates that the echoes should probe undisturbed ISM. 

To expand the argument, the ISM density varies substantially in different directions 
from Cas A (as we show below), so we would expect the wind to have penetrated
to varying depths around 50 pc. Yet there is no change in the character of 
the IR echoes. Therefore, the stellar wind is unlikely to have penetrated to 
the ISM and modified it where the IR echoes are produced.

\subsection{The Three-dimensional properties of the ambient ISM}

\subsubsection{Morphology}

Fig. 6 displays a three-dimensional picture of the distribution of dust 
surrounding Cas A. The light echoes display the structure 
of the ISM dust probed in the form of multiple sheet-like 
surfaces. Due to the interval of six months between 
successive epochs (except epoch 1 and epoch 2 with about one year time 
difference and epoch 2 and epoch 3, separated by about two months), the
sheets are separated by about 0.1 pc. 

Although the echoes are distributed all around the SNR, they
are also highly structured. We made a 
two-dimensional plot of the echoes, or dust particles, projected onto the 
plane of sky to demonstrate the asymmetrical density distribution of dust 
(see Fig. 7). We see a higher density of dust on 
the west and northwest sides of Cas A than elsewhere. 

To examine the morphology of the ISM more closely, the same small region in 
 the images at the bottom of Fig. 2 is plotted in three dimensions in Fig. 8. This figure demonstrates 
that the ISM is distributed in very well-defined filaments and sheets with 
typical dimensions larger than our resolution limit of $\sim $ 0.1 pc, but 
only slightly so in the narrowest dimensions.

The three-dimensional structure around Cas A revealed by the IR echoes 
shows the position, density, and shape of dust clouds on a scale of the 
observational limit of 0.1 pc over a region with 
projected radius of $\sim $ 50 pc (see Fig. 6). We now use this information to determine 
some general properties of the ISM surrounding Cas A. 

\subsubsection{Filling Factor}

In Fig. 9, we plot the filling factor of the small-scale ISM clumps in epoch 4 as a 
function of the projected distance, \textit{h}, within a range of 9 to 109 
parsecs and with a bin size of 10 pc, perpendicular to the line of sight. 
The SN light pulse slices the ISM in thin sheets, so the filling factor is readily
computed as the fraction of each sheet containing echoes. The 
two-dimensional coordinate system was divided into four quadrants, which 
thus yields four equal areas of the echo-generating paraboloid. In each 
quadrant, the percent ratio of the areas of echoes and of the portion of the 
paraboloid was computed within the above distance range (see Fig. 9). The 
filling factors are generally higher on the west side of Cas A, as can also be easily seen from 
Fig. 7. The filling factors tend to fall with increasing distance from the SNR. 
This overall behavior suggests that most of the dust lies behind and relatively 
near to Cas A. It is also 
possible that the attenuation of the intensity of the outgoing pulse of 
electromagnetic radiation, due to extinction as it propagates through the 
ISM, gives rise to the fainter IR light echoes at greater distances. 

The average filling factor of $\sim $ 0.4\% is overplotted as a dashed line.
This value is lower than but consistent with the filling factor, $\sim $ 1\%, determined 
by Krause et al. (2005) from the comparison of the net luminosity observed
with predictions for a flare of isotropic energy at the 
northern lobe of Cas A. Their estimation covered a length span of 6 arcsec, 
while ours covers about 40,000 square arcmin, so ours provides a more 
representative value. This result emphasizes the extreme clumpiness of the ISM at 
scales above the wavenumber \textit{k} $\sim $ 0.5 cycles pc$^{-1}$ with 
substantial portions concentrated into a very small volume. 

\subsubsection{Mass Estimation for the Warm Dust in small ISM Features}

The average spectral energy distribution (SED) of 59 selected infrared 
features in the northern lobe region of Cas A shows an increasing surface 
brightness with wavelength from 2.2 to 24$\mu$m (Krause et al. 2005). Images 
taken at 70 and 160$\mu$m exhibit no distinct echo features but display overall 
bright cirrus background emission. It is possible that the far-infrared 
(FIR) images may have cool echo features, but they are not resolvable due to 
the dominant cool diffuse background clouds. Thus, we assume that the IR 
light echoes observed at 24$\mu$m are mostly coming from warm dust grains. 
Because the SED is available over a small wavelength range, we cannot 
determine the exact blackbody energy distribution of the dust around Cas A, 
but we can assign an upper limit of $\sim $100 K to the temperature of the 
dust by taking 24$\mu$m as the peak wavelength. 

The total flux of the IR light echoes was computed and used to deduce the 
mass of the heated dust within successive parabolic volumes defined by 
epochs 4 and 5. The dust mass is determined from the 24$\mu$m emission by 
(Marleau et al. 2006):

\begin{equation}
\label{eq5}
{M_{d} = \frac{S_{\nu}  D^{2}}{\kappa _{d} B_{\left( {\nu ,T} \right)}}} ,
\end{equation}

\noindent
where ${S_{\nu} } $ is the total flux of echoes from epoch 4 and epoch 5, 
\textit{D} is the distance to the dust, ${\kappa _{d}} $ is the mass 
absorption opacity of the dust, for which a value of 492 
cm$^{2}$g$^{-1}$ at 24$\mu$m is assumed (Li \& Draine 2001), and ${B_{\left( 
{\nu ,T} \right)}} $ is the Planck function. The total mass 
estimate for dust within the volume of the small echo features is $\sim $ 
0.014 M$_\odot$, yielding a gas mass of $\sim $ 1.4 M$_\odot$ for a 
gas-to-dust mass ratio of 100. Since this estimate is based on an upper limit
for the dust temperature, the mass is a lower limit. The full volume occupied by the echoes
and surrounding space for this determination is $3 \times 10^5 ~ pc^3$, and the
average (lower limit) ISM density represented by the echoes is therefore only $3 \times 10^{-25}$
kg m$^{-3}$.

We have tested whether the estimated mass is affected by stochastic heating. 
We ran heating simulations for the SN echo using silicate grains. For these 
runs, we assumed that the echo was the result of the SN explosion, that the 
SN had a bolometric luminosity of 5.0 $\times$ 10$^{9}$ L$_\odot$ (Young et 
al. 2006), and that the echoes are at a distance of 67 pc from the SNR. We 
adopted the SED of SN 1987A at day 100 (Arnett et al. 1989). The grain 
sizes and their heating and cooling times are listed in Table 1. The heating 
time is the average interval between photon absorptions.

As can be seen in Table 1, even at 20\AA ~the grain will usually be 
hit by a photon with substantial energy well before it can cool significantly. The only 
grains that will show some stochastic heating will be $\le $ 15\AA ~in size. 
Other grain types (e.g., graphite) do not differ
in heat capacity and radiative properties sufficiently to modify the size range
substantially (e.g., Dwek 1986). Such small  
grains will not emit efficiently at 24$\mu$m. Thus, it is very unlikely for the 
grains around Cas A to be heated stochastically to a sufficient extent to 
undermine our mass estimate. 

\subsubsection{The Fractal Nature of the ISM and the Total Mass Estimation}

There is a growing consensus that turbulence plays a major role in the 
dynamics of molecular clouds and results in fractal structures (Falgarone \& 
Phillips 1990; Elmegreen \& Falgarone 1996). Fractals are 
fragmented geometric shapes that show self-similarity at all levels of 
magnification (Mandelbrot 1983). 

To test the fractal nature of the structures, we selected two 
echo-dominant regions observed with a time interval of one year. We computed
the two-dimensional power spectrum of the difference image (Fig. 3). Since it is 
roughly symmetric around the central peak, we converted it to a one-dimensional
power spectrum by azimuthal averaging. We used 
several subregions of the {\it Spitzer} mosaics with no detectable emission as 
noise images, and analyzed them in the same way. The individual power 
spectra from the noise images were 
summed and averaged, and the averaged noise power spectrum was subtracted 
from the power spectrum of the original image. Since the difference image of adjacent
epochs isolates a thin layer of the ISM from the 
full depth of the ISM surrounding the SNR, the resulting power spectrum
minimizes effects due to the superposition of regions along the line of sight.   

The power spectra (original, noise, and noise-subtracted) of two sample regions are 
compared in Fig. 10. The power spectrum from the first region displays a slight 
transition of the slope at the frequency of $\sim $ 2.0 cycles pc$^{-1}$, most 
likely due to the decrease in intensity of the components of the features 
that are almost invariant over the time difference of one year. That 
assumption becomes more certain because the power spectrum from the second region 
shows less of a distinct transition of the slope. The second region is located 
further from the line of sight to the SNR than the first; thus, the spatial 
change in the time difference in the second region results in a greater apparent 
motion in two dimensional images. For both regions, the 
behavior of the noise-subtracted data is reasonably consistent with a power 
law, k$^{-\beta}$ with an index of $\beta$ = 3.5. This slope was found by Ingalls et 
al. (2004) for a 24$\mu$m image of the general ISM. This behavior is as expected 
for fractal structure. That is, our three-dimensional images confirm the 
fractal structure of the ISM previously deduced from studies in only two 
dimensions. 

Assuming that the power spectrum of the two regions is described by the 
power-law behavior with the spectral index 3.5 over the low frequency range 
from $\sim $ 10$^{-3}$ to 1.0 cycles pc$^{-1}$ (as indicated by Ingalls et al. 
(2004) in the wavenumber range from $\sim $ 10$^{-3}$ to 10$^{-1}$), we 
estimated the total mass in the ISM from $\sim $ 0.01 cycles pc$^{-1}$  to $\sim $ 
5 cycles pc$^{-1}$. The result is $\sim $ 6.2 $\times$ 10$^{-23}$ kg m$^{-3} $, 
about 200 times more than in the small (k $>$ 0.5 cycles pc$^{-1}$) features alone. The
derived total density is slightly 
higher than the average mass density of dust in the diffuse ISM of $\sim $ 1.8 $\times$ 
10$^{-23}$ kg m$^{-3}$ (Whittet 2003). Given the uncertainties, particularly in the
dust temperature, this value agrees well with expectations (e.g., \S 5.1). A corollary of the modest
density associated with the echoes is that they are optically thin to
the supernova radiation and that we are measuring volume emission, not just
that from a surface layer.

\subsubsection{Comparison with Previous Work}

Complexes of filamentary and wavy knots have been found in many 
observations of the small-scale ISM environment. Examples include the 
Chamaeleon III region (Gahm et al. 2002), elephant trunk structures (Carlqvist et 
al. 1998), the Lagoon nebula (Caulet 1997), diffuse molecular clouds 
(Falgarone et al. 2004), and a nearby cold interstellar cloud 
(McClure-Griffiths et al. 2006). Typically, ISM filaments are distributed in 
long, finger-like patterns and are observed overlapping and crossing each 
other due to projection effects. These overlapping filaments can be 
identified by differences in radial velocity and in color intensity at the 
observed wavelength, e.g., the contrast between dark and bright filaments.

The individual elongated filamental features in Fig. 7 are reminiscent of 
those studied in two dark clouds, the Lynds 204 complex and the Sandqvist 
187-188 complex, by Carlqvist and Gahm (1992). They suggested that the 
Bennett pinch, resulting from a cylindrical plasma carrying an axially 
directed current, could be the physical mechanism for formation of the wavy 
filaments. In this process, the plasma undergoes a pinch compression by the 
current simultaneously with the generation of a toroidal magnetic field. 
Verschuur (1995) has reviewed structures observed with high resolution HI 
data. He also suggests the pinching action of electric currents as 
contributing to the highly filamentary structures of the ISM. Detailed 
numerical models invoking such processes have been described by Fiege and 
Pudritz (2000).

It is not clear whether the filaments are stable or unstable. The 
generalized Bennett relationship for the equilibrium of a filament consists 
of two inward pinching forces, due to the current and to gravity, and three 
outward forces due to gas pressure, to the axial magnetic field, and to 
rotational energy around the filament axes (Verschuur 1995). There is a 
possibility that the pinching action undergoes various kinds of 
instabilities and develops dynamically. Carlqvist and Gahm (1992) discuss 
two such examples: 1.) kink instability, when motions in a conduction fluid 
body change the initial magnetic field in a way that it becomes unstable and 
produces a kink (Alfv\'en and F\"althammar 1963); and 2.) sausage instability, 
which originates from a small radial disturbance of a pinched filament and 
may lead to disruption of the filament. Comparisons of the wavy shapes of 
the filaments in our data with the predictions of such models can be used to 
test theoretical hypotheses for filament formation and maintenance. 

Molecular clouds have also been described as fractal and highly dynamical 
(Falgarone et al. 2004). Fractal structures are characterized by a power law 
power spectrum of the image ($ \propto $ k$^{-\beta} $) and a completely random 
distribution of the image phases (Stutzki et al. 1998). Several studies have 
obtained power spectra of the ISM that show the characteristic fractal 
self-similar structure, with the spectral indices $\beta$ $\sim $ 2.6 to 2.8 
(Green 1993) or $\beta$ $\sim $ 2.6 to 3.5 (Ingalls et al. 2004). Stutzki et al. 
(1998) introduced the 2-dimensional $\Delta$ - variance method, which has been used 
to find $\beta$ $\sim $ 2.5 (Bensch et al. 2001; Falgarone et al. 2004, and 
references therein). We show that the power spectrum found by Ingalls et al. 
(2004) with $\beta$ = 3.5 is consistent with the three dimensional ISM structure, 
and we can place a physical scale on the structures where we have fitted it. 

\section{Conclusion}

IR light echo observations at 24$\mu$m have allowed the exposition of the 
three-dimensional physical structure of the ISM around the SNR, Cas A. From 
high pass filtering to subtraction, we have shown the detailed analysis 
required to distinguish as many IR light echoes as possible from any diffuse 
background and other fixed sources, such as stars and bright nebular 
emission. 

From our results, we draw the following main conclusions:

$ \bullet $ The detailed behavior of the IR light echoes confirms that they 
are re-radiation of the absorbed 
outward-moving SN light pulse by the ISM surrounding Cas A.

$ \bullet $ The three-dimensional display of the echoes shows different 
sizes of clumps connected with filaments that trace sheet-like patterns. 

$ \bullet $ The surface filling factor of the heated dust grains is about 
0.4\%, for fine structures (spatial frequencies $\ge$ 0.5 cycles pc$^{-1}$).

$ \bullet $ The structures have a power-law power spectrum k$^{-\beta}$ with index $\beta$ $\sim 
$ 3.5, consistent with fractal structure.

$ \bullet $ The average ISM density above 0.5 cycles pc$^{-1}$ and below 5 cycles pc$^{-1}$
in the region we have studied is $\sim 3 \times 10^{-25}$ kg m$^{-3}$.  
If we integrate the power spectrum down to $\sim $ 0.01 cycles pc$^{-1}$, 
we estimate that the ISM surrounding Cas A is $\sim 6 \times 10^{-23}$ kg m$^{-3}$, 
more dense than the general diffuse ISM by a factor of a few.

\acknowledgments

This work is based on observations made with the {\it Spitzer} Space 
Telescope, which is operated by the Jet Propulsion Laboratory, California 
Institute of Technology under contract with NASA. Support for this work 
was provided by NASA through JPL/Caltech contracts 1276396 and 1255094 to the University of Arizona.

\clearpage

\begin{deluxetable}{ccc}
\tabletypesize{\scriptsize}
\tablecaption{Silicate dust heating/cooling \label{data_table}}
\tablewidth{0pt}
\tablehead{
\colhead{Grain Size} & \colhead{Heating Time}& \colhead{Cooling Time} \\
\colhead{(\AA)}  & \colhead{(sec)}& \colhead{(sec)}
}
\startdata

10          & 381   & 12      \\
20          & 50    & 125     \\
30          & 14    & 435     \\
40          & 6     & 1041    \\

\enddata

\end{deluxetable}

\clearpage
\begin{figure}
\epsscale{1.0}
\plotone{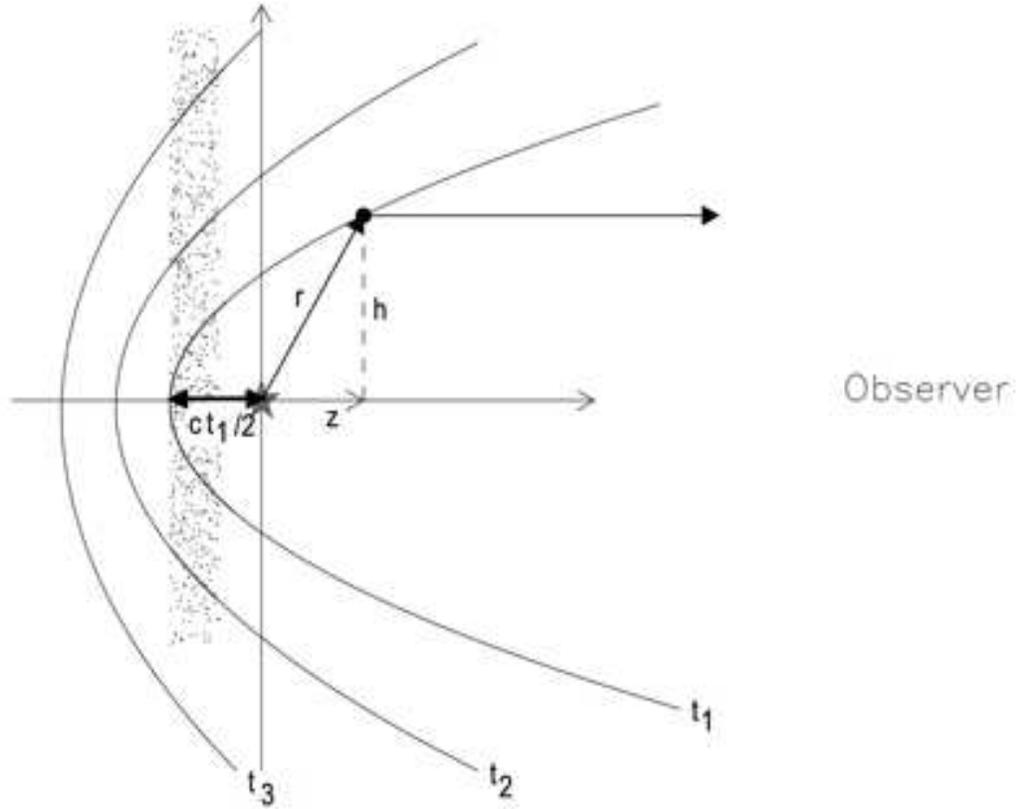}
\caption{Geometry of the IR light echo propagation. The red star represents 
Cas A. Assuming a dust column behind Cas A, the light-red colored region in 
each parabola (a two-dimensional parabolic surface parallel to the line of 
sight) defines the dust heated by light from the initial outburst at a 
specific time, t$_1$ $<$ t$_2$ $<$ t$_3$, since the supernova outburst. This heated dust 
produces the IR echoes. 
\label{01}}
\end{figure}
\clearpage

\begin{figure}
\epsscale{.8}
\plotone{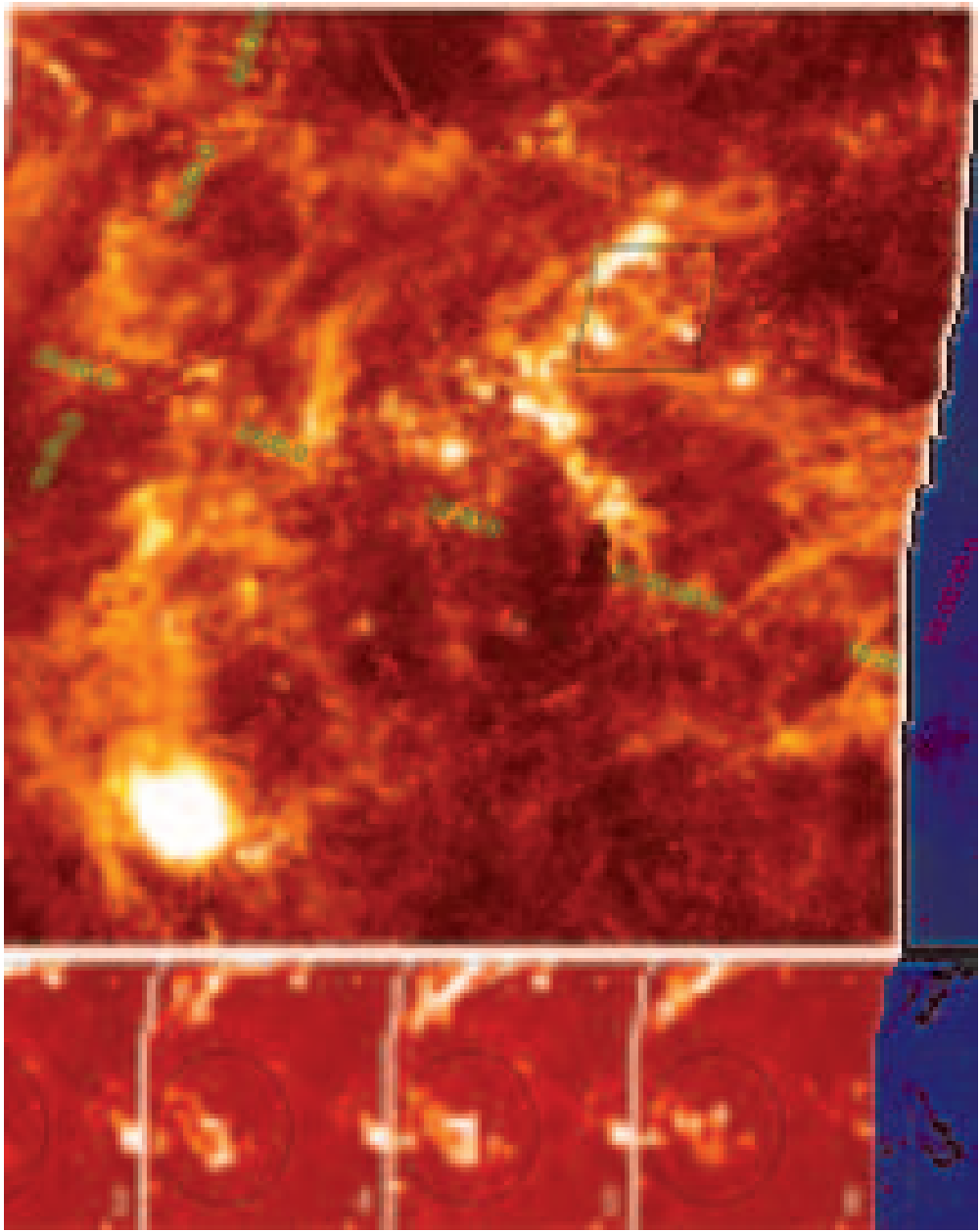}
\caption{24$\mu$m images of moving features (at the bottom) about 45 pc in 
projected distance from Cas A in the northern region (the subregion shown in 
a square box) . Cas A is the big bright blob on the left lower side of the 
top image. The images were obtained on 2005 February 3 (epoch 3), 2005 September 
1 (epoch 4), 2006 February 14 (epoch 5), and 2006 October 2 (epoch 6). 
Nearly all the bright features in this region are IR light echoes, and they 
are especially evident in the circles.
\label{02}}
\end{figure}
\clearpage

\begin{figure}
\epsscale{1.0}
\plotone{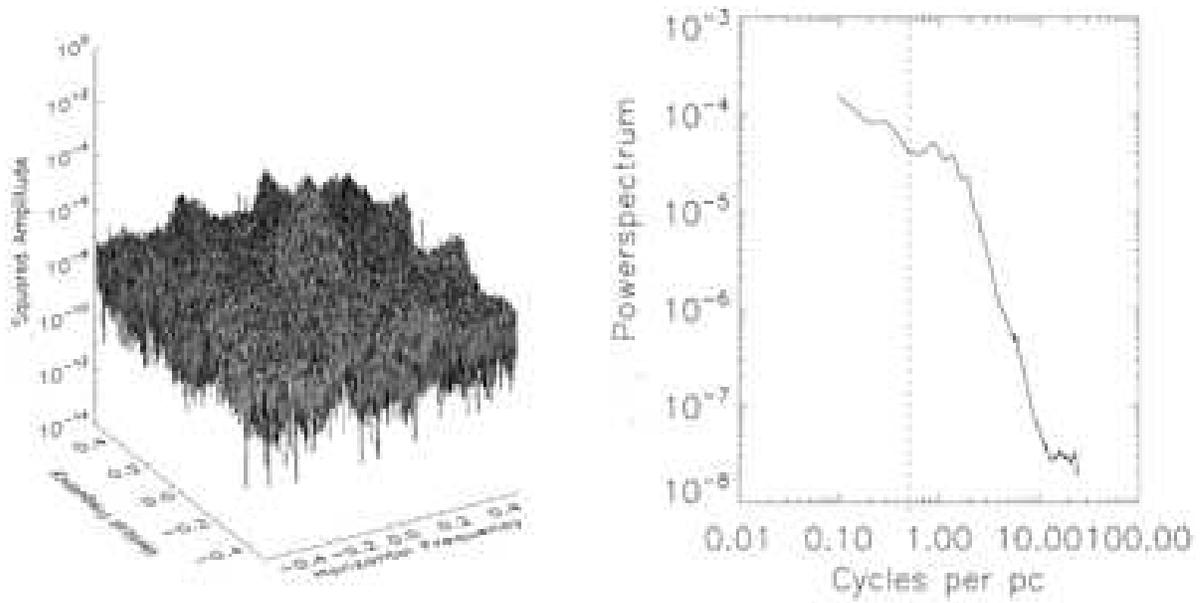}
\caption{Surface plot of the power spectrum of the difference images (mostly IR 
light echoes, Cas A, bright stars, and nebulae) on the left and the averaged 
power spectrum in log scale on the right. The cutoff 
frequency, where the filter is half its maximum, is defined by the vertical
dashed line. 
\label{03}}
\end{figure}
\clearpage

\begin{figure}
\epsscale{1.0}
\plotone{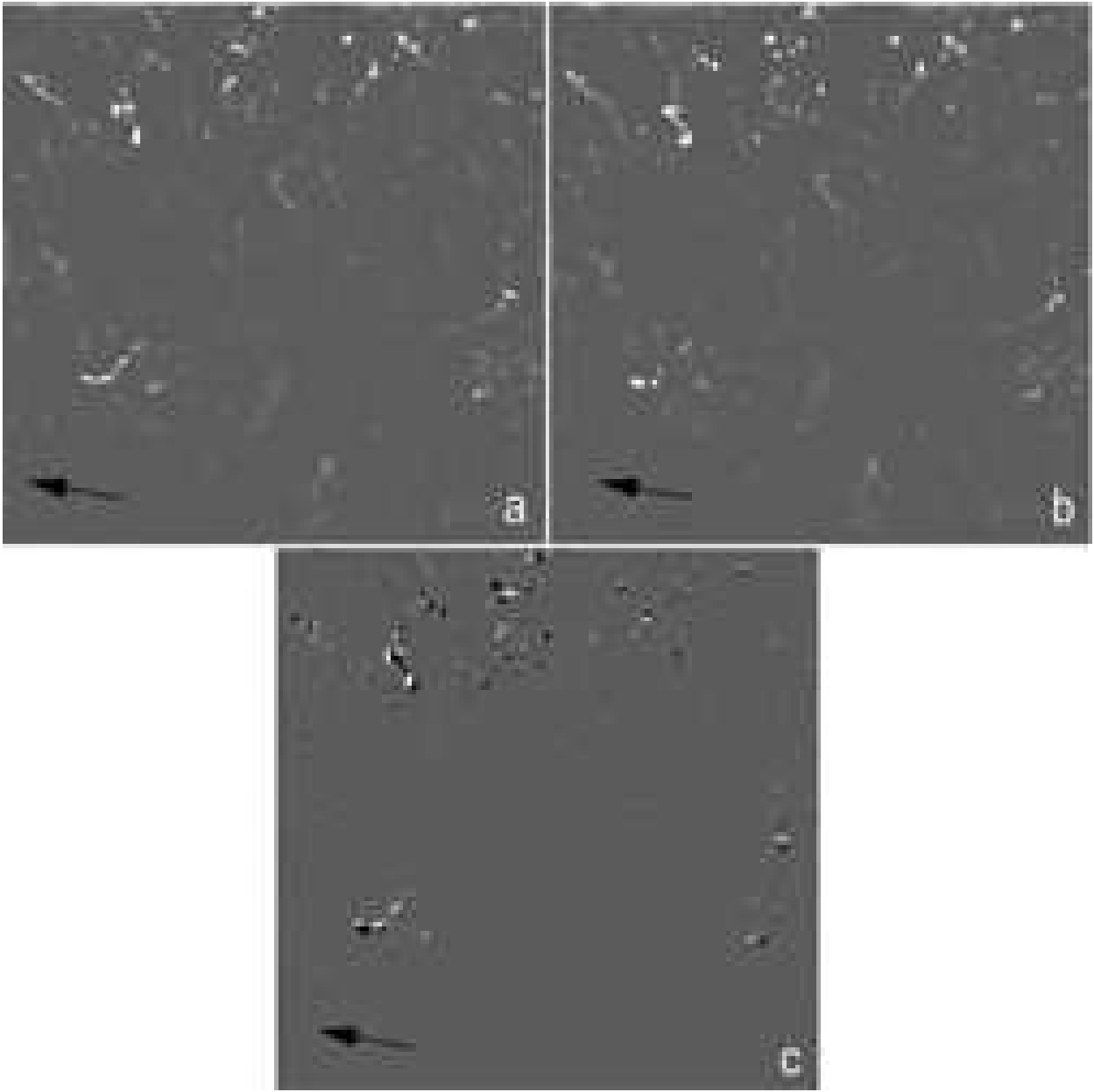}
\caption{Subtraction of two different epoch images. Image (a) is a 
small region extracted from the epoch 4 image, while image (b) is 
from the epoch 5 image. The difference image is shown in (c). The 
bright white regions belong to epoch 4 while the black regions belong to 
epoch 5. The location of Cas A with respect to the subregions is indicated by 
arrows.
\label{04}}
\end{figure}
\clearpage

\begin{figure}
\epsscale{1.0}
\plotone{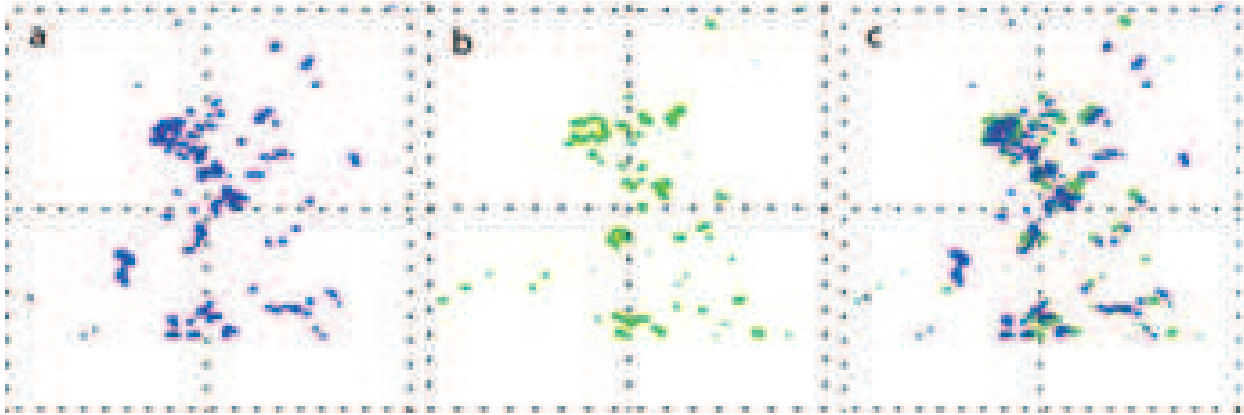}
\caption{Detected pattern of ISM material for a 20pc $\times$ 20pc region: a)
is epoch 1; b) is epoch 2; and c) is their combination.  
\label{05}}
\end{figure}
\clearpage

\begin{figure}
\epsscale{1.0}
\plotone{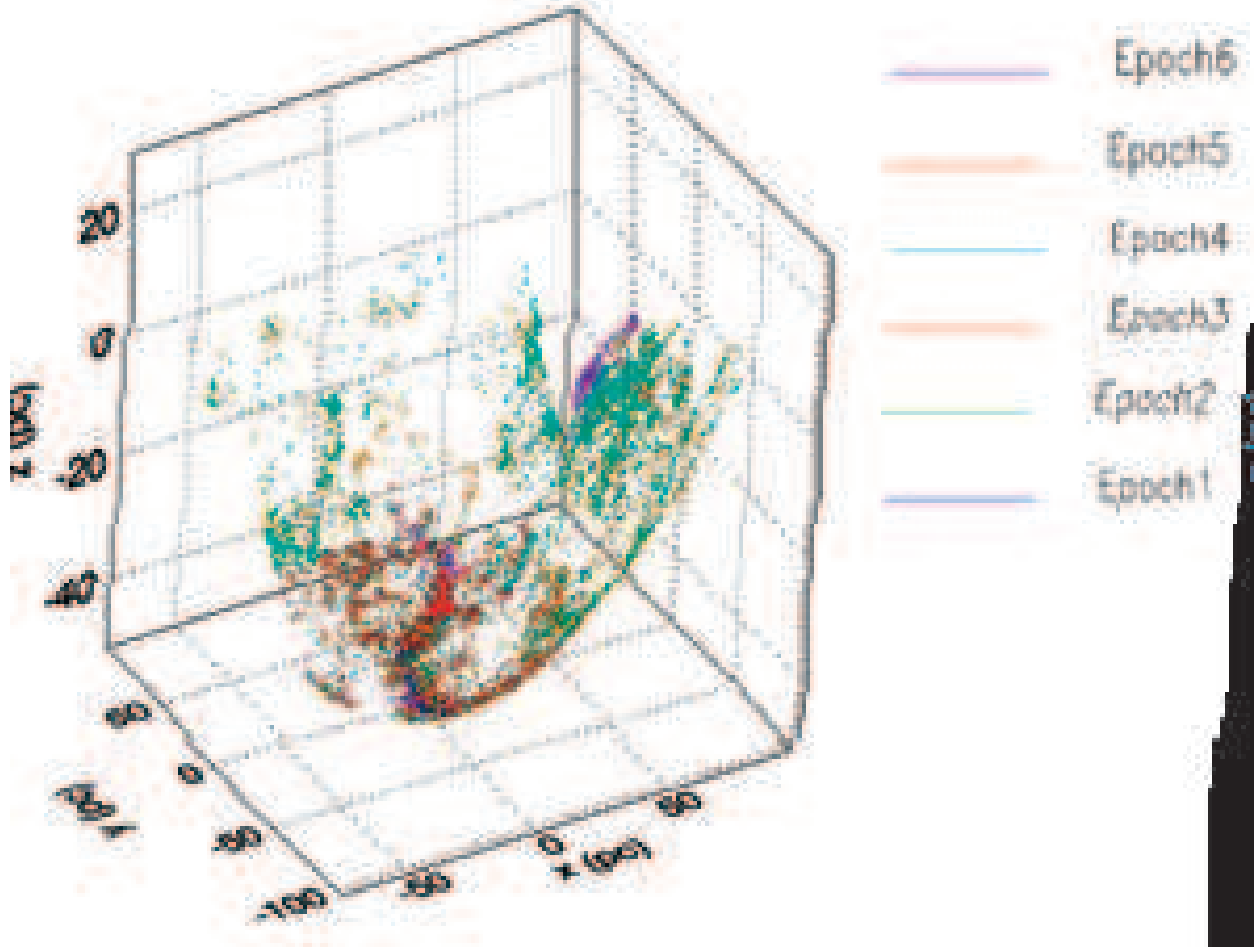}
\caption{A three-dimensional plot of dust from all six epochs over about three 
years. Cas A is located at the origin (0,0,0), and \textit{z} values 
increase toward the observer on Earth. The unit in \textit{x}, \textit{y}, 
and \textit{z} is parsecs. 
\label{06}}
\end{figure}
\clearpage

\begin{figure}
\epsscale{1.0}
\plotone{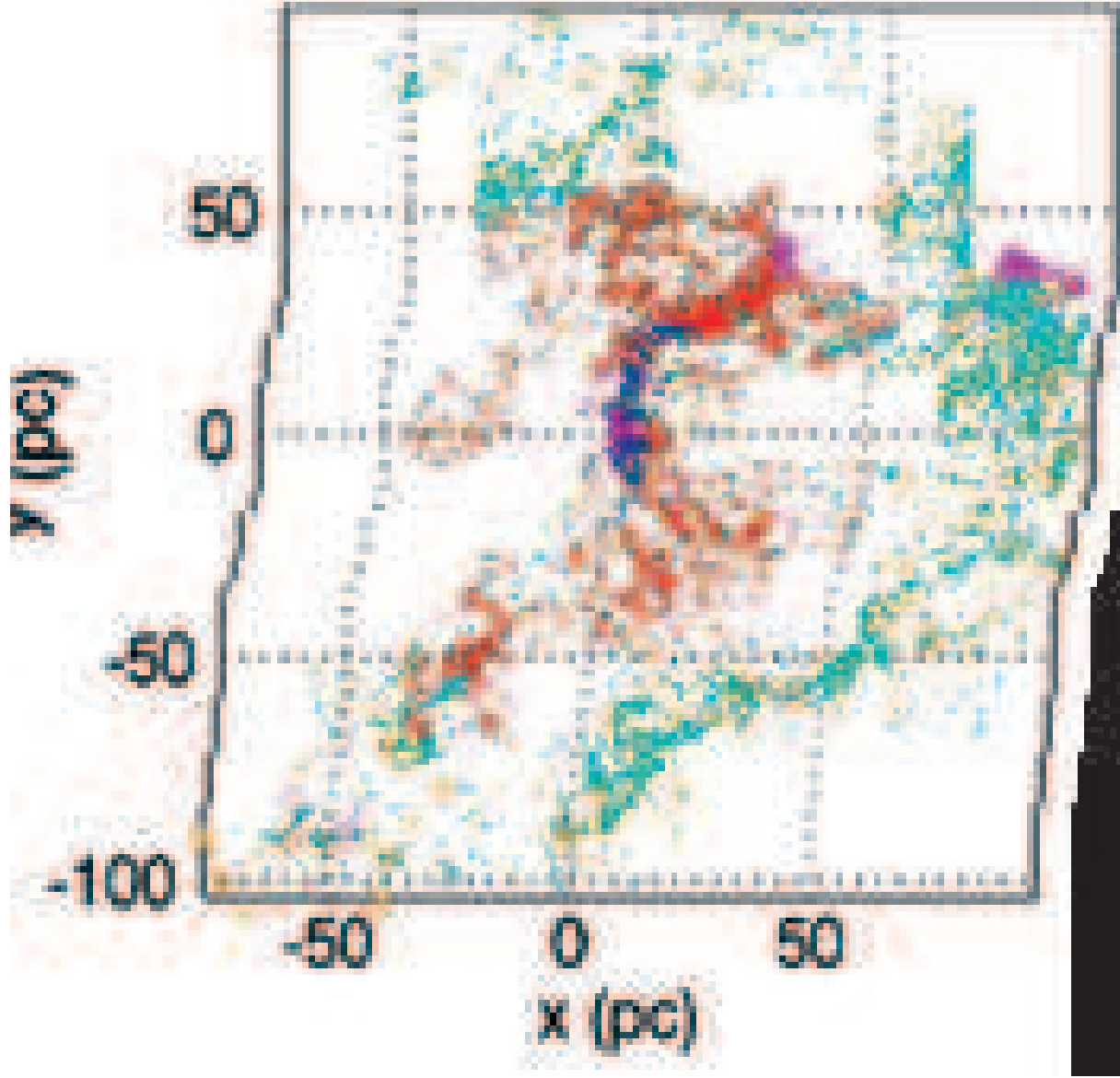}
\caption{Projected image of the three-dimensional plot as viewed by a distant 
observer. Cas A is located at the origin (0,0). The same color 
representations as in Fig. 6 are used.
\label{07}}
\end{figure}
\clearpage

\begin{figure}
\epsscale{1.0}
\plotone{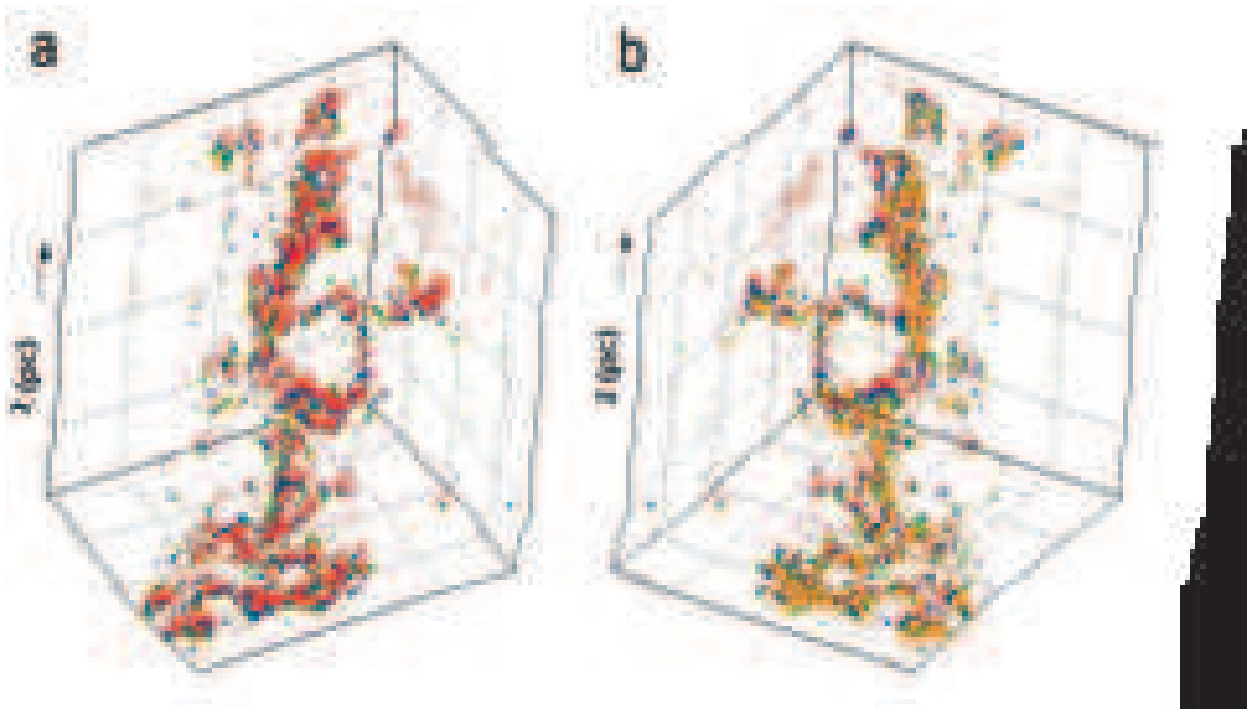}
\caption{The two images show the three-dimensional structure of the ISM dust 
in the northern region of Cas A, from both sides. The dotted grid indicates a spacing of 2 pc
in the z axis (toward the observer) and 5 pc in the x and y axes. The colors are 
red in epoch 3, blue in epoch 4, green in epoch 5, and orange in epoch 6. 
\label{08}}
\end{figure}
\clearpage

\begin{figure}
\epsscale{1.0}
\plotone{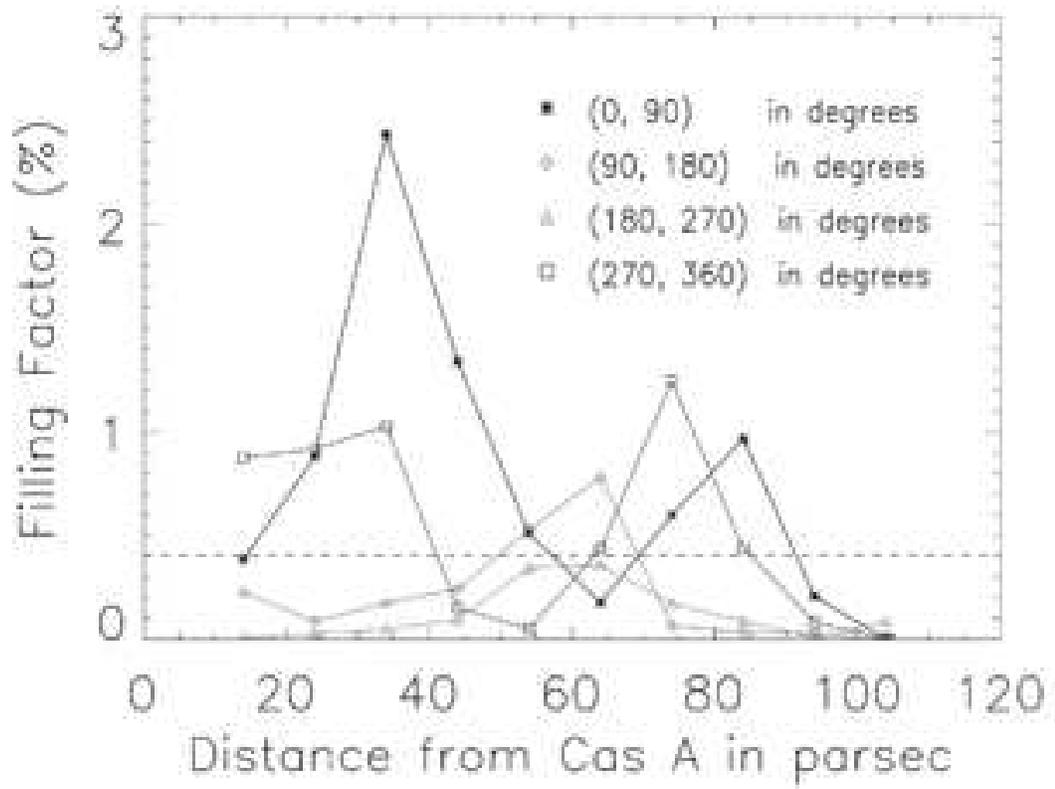}
\caption{Filling factors in percent as a function of projected distance, {\it h}, in four different angular 
quadrants in epoch 4. The dashed line represents the average filling factor, 0.4\%. 
\label{09}}
\end{figure}
\clearpage

\begin{figure}
\epsscale{1.0}
\plotone{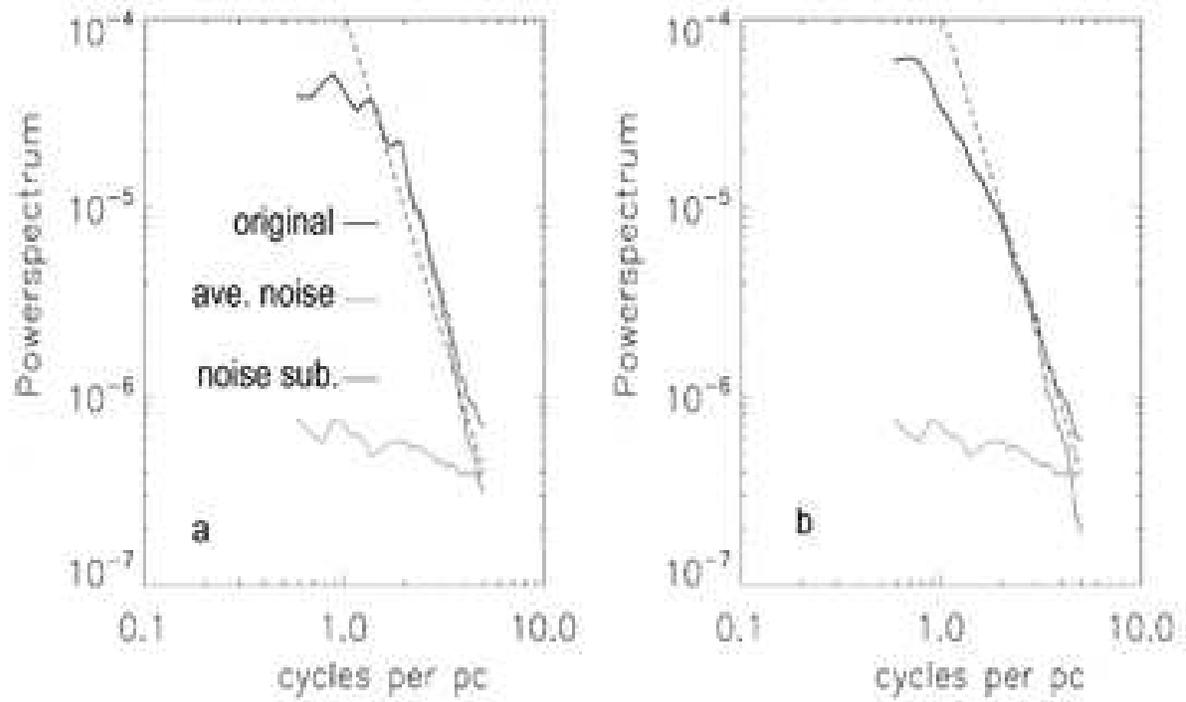}
\caption{Power spectra of two sample regions.The lines are coded for both
images as indicated in (a): 
original data (black solid line), the noise (red solid line), and 
noise-subtracted data (blue solid line). The dashed line represents a power law 
with a spectral index of 3.5. 
\label{10}}
\end{figure}


\begin{thebibliography}{}

\bibitem[Alfv\'en, H, et al. 1963]{Alf63}Alfv\'en \& F\"althammar, C.-G. 1963, Cosmical Electrodynamics (2nd ed.; 
Oxford: Carendon)
\bibitem[Arnett, et al. (19890]{arn89}Arnett, W. D., Bahcall, J. N., Kirshner, R. P., \& Woosley, S. E. 1989, 
ARA\&A, 27, 629
\bibitem[Bensch, et al. (2001)]{ben01}Bensch, F., Stutzki, J., \& Ossenkopf, V. 2001, A\&A, 366, 636
\bibitem[Boss (2005)]{bos05}Boss, A. P. 2005, ApJ, 622, 393
\bibitem[Braun, et al.(2005)]{bra05}Braun, R., \& Kanekar, 2005, A\&A, 436L, 53
\bibitem[Bruzual, et al.(1988)]{bru88}Bruzual, A. G., Magris, C. G., \& Calvert, N. 1988, ApJ, 333, 673
\bibitem[Burton et al. (1990)]{bur90}Burton, M. G., Hollenbach, D. J., \& Tielens, A. G. M. 1990, ApJ, 365, 620
\bibitem[Carlqvist, et al.(1992)]{car92}Carlqvist, P., \& Gahm, G. F. 1992, IEEE Trans. Plasma Sci. 20, 867
\bibitem[Carlqvist, et al.(1998)]{car98}Carlqvist, P., Kristen, H., \& Gahm, G. F. 1998, A\&A, 332, L5
\bibitem[Caulet (1997)]{cau97}Caulet A. 1997, ST-ECF Newslett., 24, 12
\bibitem[Draine, et al.(2003)]{dra03}Draine, B. T. 2003, ARAA, 41, 241
\bibitem[Dwarkadas (2004)]{dwa04}Dwarkadas, V. 2004, in Cambridge contemporary astrophysics ser., Cosmic 
explosions in three dimensions : asymmetries in supernovae and gamma-ray bursts, ed. P. Hoflich, 
P. Kumar and J. C. Wheeler, (Cambridge, UK: Cambridge University Press), 74
\bibitem[Dwek (1986)]{dwe86}Dwek, E. 1986, ApJ, 302, 363
\bibitem[Elmegreen \& Falgarone (1996)]{elm96}Elmegreen, B. G., \& Falgarone, E. 1996, ApJ, 471, 816
\bibitem[Falgarone \& Phillips (1990)]{fal90}Falgarone, E. \& Phillips, T. G. 1990, ApJ, 359, 344
\bibitem[Falgarone et al. (2004)]{fal04}Falgarone, E., Hily-Blant, P., \& Levrier, F. 2004, Ap\&SS, 292, 89
\bibitem[Fesen (2001)]{fes01}Fesen, R. A. 2001, ApJ, 133, 161
\bibitem[Fiege \& Pudritz (2000)]{fie00}Fiege, J. D., \& Pudritz, R. E. 2000, MNRAS, 311, 85
\bibitem[Fischera \& Dopite (2005)]{fis05}Fischera, J. \& Dopita, M. 2005, ApJ, 619, 340
\bibitem[Gahm et al. (2002)]{gah02}Gahm, G. F., Lehtine, K., Carlqvist, P., Harju, J., Juvela, M, \& Mattila, 
K. 2002, A\&A, 389, 577
\bibitem[Garcia et al.(1996)]{gar 96}Garc\'ia-Segura, G., Langer, N., \& Mac Low, M.-M. 1996, A\&A, 316, 133
\bibitem[Gordon et al. (2005)]{gor05}Gordon, K. D., et al. 2005, PASP, 117, 503
\bibitem[Green (1993)]{gre93}Green, D.A. 1993, MNRAS, 262, 327
\bibitem[Heger et al.(1997)]{heg97}Heger, A., Jeannin, L., Langer, N., \& Baraffe, I. 1997, A\&A, 327, 224
\bibitem[Hagmann \& Kegel (2003)]{heg03}Hegmann, M. \& Kegel, W. H. 2003, MNRAS, 342, 453
\bibitem[Indebetouw et al. (2006)]{ind06}Indebetouw, R., Whitney, B. A., Johnson, K. E., \& Wood, K. 2006, ApJ, 636, 362
\bibitem[Ingalls, et al.(2004)]{ing04}Ingalls, J. G., et al. 2004, ApJSS, 154, 281
\bibitem[Inoue (2005)]{ino05}Inoue, A. K. 2005, MNRAS, 359, 171
\bibitem[Kim \& Martin (1996)]{kim96}Kim, S.-H., \& Martin, P. G. 1996, ApJ, 462, 296
\bibitem[Krause, et al.(2004)]{kra04}Krause, O., Birkmann, S. M., Rieke, G. H., Lemke, D., Klaas, U., Hines, D. 
C., \& Gordon, K. D. 2004, Nature, 432, 596
\bibitem[Krause et al. (2005)]{kra05}Krause, O., et al. 2005, Science, 308, 1604
\bibitem[Li \& Draine (2001)]{li01}Li, A., \& Draine, B. T. 2001, ApJ, 554, 778
\bibitem[Mandelbrot (1983)]{man83}Mandelbrot, B. B., 1983, The Fractal Geometry of Nature, Freeman, San 
Francisco.
\bibitem[Marleau et al. (2006)]{mar06}Marleau, F. R., et al. 2006, ApJ, 646, 929
\bibitem[McClure-Griffiths et al. (2006)]{mcc06}McClure-Griffiths, N. M., Dickey, J. M., 
Gaensler, B. M., Green, A. J., \& Haverkorn, M. 2006, ApJ, 652, 1339
\bibitem[Patat (2005)]{pat05}Patat, F. 2005, MNRAS, 357, 1161
\bibitem[Rieke et al. (2004)]{rie04}Rieke, G. H., et al. 2004, ApJS, 154, 25
\bibitem[Russell (2003)]{rus03}Russeil, D. 2003, A\&A, 397, 133
\bibitem[Smith, et al.(2007)]{smiliu07}Smith, J. D. T., et al. 2007, ApJ, 656, 770
\bibitem[Stinebring et al. (2000)]{sti00}Stinebring, D. R., Smirnova, T. V., Hankins, T. H., Hovis, J. S.,
Kaspi, V. M., Kempner, J. C., Myers, E., \& Nice, D. J. 2000, ApJ, 539, 300
\bibitem[Stutzki et al.(1998)]{stu98}Stutzki, J., Bensch, F., Heithausen, A., Ossenkopf, V., Zielinsky, M., 1998, 
A\&A, 336, 697
\bibitem[Thorstensen et al. (2001)]{tho01}Thorstensen, J. R., Fesen, R. A., \& van den Bergh, S. 2001, ApJ, 122, 297
\bibitem[V\'azquez-Semadeni, et al. (2007)]{vaz07}V\'azquez-Semadeni, E., G\'omez, G.C., Jappsen, A.-K., 
Ballesteros-Paredes, J., Gonz\'alez, R.F., Klessen, R.S., 2007, ApJ, 657, 870
\bibitem[Verschuur(1995)]{ver95}Verschuur, G. 1995, Ap\&SS. 227, 187
\bibitem[Whittet (1989)]{whi89}Whittet, D. C. B. 1989, "The composition of dust in stellar ejecta," in
Allanmandola, L. J., \& Tielens, A. G. G. M. (eds), Interstellar Dust, (Dordrecht:Kluwer), pp 455-466
\bibitem[Whittet (2003)]{whi03}Whittet, D. C. B. 2003, Dust in the Galactic Environment (2d ed.; Bristol: 
IoP)
\bibitem[Wilson et al. (1993)]{wil93}Wilson, T. L., Mauersberger, R., Muders, D., Przewodnik, A., \& Olano, C. A. 
1993, A\&A, 280, 221
\bibitem[Witt \& Gordon (19964)]{wit96}Witt, A. N., \& Gordon K. D. 1996, ApJ, 463, 681
\bibitem[Wolf et al. (1999)]{wol99}Wolf, S., Henning, Th., \& Stecklum, B. 1999, A\&A, 349, 839
\bibitem[Xu, et al.(1995)]{xu95}Xu, J., Crotts, A. P. S., \& Kunkel, W. E. 1995, ApJ, 451, 806
\bibitem[Wood et al. (2005)]{woo05}Wood, K., Haffner, L. M., Reynolds, R. J., Mathis, J. S., \& 
Madsen, G. 2005, ApJ, 633, 295
\bibitem[Young, et al.(2006)]{you06}Young, P. A., et al. 2006, ApJ, 640, 891
\end{thebibliography}
\end{document}